\documentstyle[12pt]{article}
\catcode`\@=11
\def\eqnarray{\let\@currentlabel=\theequation\refstepcounter{equation}
    \global\@eqnswtrue
    \global\@eqcnt\z@\tabskip\@centering\let\\=\@eqncr
    $$\halign to \displaywidth\bgroup\@eqnsel\hskip\@centering
      $\displaystyle\tabskip\z@{##}$&\global\@eqcnt\@ne
       \hfil${{}##{}}$\hfil
      &\global\@eqcnt\tw@ $\displaystyle\tabskip\z@{##}$\hfil
       \tabskip\@centering&\llap{##}\tabskip\z@\cr}
\def\lefteqn#1{\hbox to 4\arraycolsep{$\displaystyle #1$\hss}}
\newcommand{\beq}{\begin{equation}}
\newcommand{\eeq}{\end{equation}}
\def\IR{{\hbox{{\rm I}\kern-.2em\hbox{\rm R}}}}
\def\rref#1{(\ref{#1})}
\begin{document}
\begin{flushright}    
UCD-96-10\\           
gr-qc/9603049\\       
March 1996\\          
\end{flushright}      
\vspace{1ex}         
\begin{center}
{\bf
THE STATISTICAL MECHANICS OF HORIZONS\\
AND BLACK HOLE THERMODYNAMICS}\footnote{Talk given at the Pacific
Conference on Gravitation and Cosmology, Seoul}
\vskip .75cm      
S.\ CARLIP\footnote{e-mail: carlip@dirac.ucdavis.edu}
\vskip .75cm      
{\it
Department of Physics\\
University of California at Davis, \\
Davis, CA 95616, U.S.A.
}
\end{center}
\vskip .7cm      
\noindent {\bf Abstract}
\vskip 0.7cm
Although we know that black holes are characterized by a temperature
and an entropy, we do not yet have a satisfactory microscopic
``statistical mechanical'' explanation for black hole thermodynamics.
I describe a new approach that attributes the thermodynamic properties
to ``would-be gauge'' degrees of freedom that become dynamical
on the horizon.  For the (2+1)-dimensional black hole, this approach
gives the correct entropy.
\vskip 1cm

\noindent {\bf I. Introduction}
\vskip 0.7cm
It has now been more than twenty years since Bekenstein \cite{Bek}
and Hawking \cite{Hawking} demonstrated that black holes are
thermodynamic objects, with characteristic temperatures and
entropies.  The evidence for black hole thermodynamics is convincing:
the same temperatures and entropies can be obtained from a wide variety
of approaches, ranging from the study of quantum field theory in black
hole backgrounds \cite{Hawking,Waldbk} to semiclassical path integration
\cite{Gibbons,BrownYork} to appeals to the consistency of standard
thermodynamics in the presence of black holes \cite{Bek,Wald}.  But
despite considerable effort, we do not yet have a satisfactory
``statistical mechanics'' of black hole thermodynamics; we cannot
explain the temperature and entropy in terms of microscopic degrees
of freedom.  The entropy of the Universe is, apparently,
\begin{eqnarray}
S &=&
  {1\over4}\sum\, (\hbox{areas of black hole horizons}) \nonumber\\[.1ex]
  &\phantom=& + \sum\, (\hbox{statistical mechanical entropies of
  everything else}) .
\label{a1}
\end{eqnarray}
This lack of symmetry is disturbing.

Let us suppose that black hole thermodynamics has an undiscovered
statistical mechanical origin.  It is reasonable to expect that the missing
microscopic degrees of freedom should come from quantum gravity---after
all, the relevant thermodynamic quantities can be obtained from a path
integral for gravity with no additional couplings.  But (2+1)-dimensional
gravity then presents a paradox.  Ba{\~n}ados, Teitelboim, and Zanelli
showed in 1992 that general relativity in three spacetime dimensions admits
black hole solutions \cite{BTZ}, and these black holes have thermodynamic
properties not unlike those in 3+1 dimensions \cite{CarTeit}.  But the
physical degrees of freedom of (2+1)-dimensional gravity are fairly well
understood (see, for example, \cite{Car1}), and it is easy to see that
there are simply not enough degrees of freedom in the conventional
formulation to account for the predicted entropy.  Something is missing.

The goal of this article is to describe a new approach, discovered
independently by Balachandran, Chandar, and Momen \cite{Bal1,Bal2} and
by me \cite{Car2,Car3}, that may provide the missing degrees of freedom.
The basic argument is that in the presence of a black hole horizon,
certain field excitations that are normally discarded as ``pure gauge''
become physical.  In 2+1 dimensions, these degrees of freedom may be
counted, and given some reasonable assumptions about quantization,
they yield the correct entropy.  A (3+1)-dimensional version of this
counting argument does not yet exist, but work is in progress.
\vskip 1cm

\noindent {\bf II. Gauge-Fixing, Boundaries, and Physical Degrees of
Freedom}
\vskip 0.7cm
General relativity has a large gauge group, the group of spacetime
diffeomorphisms, and the number of physical degrees of freedom is
correspondingly smaller than it first appears.  Let us briefly
review two methods for identifying the physical degrees of freedom:
\vskip 0.2cm
\noindent {\bf 1.\ The York splitting} \cite{York}: In an $(n+1)$-dimensional
spacetime, let $g_{ij}$ denote the induced metric on an $n$-dimensional
spacelike hypersurface $\Sigma$.  Any small fluctuation $\delta g_{ij}$
of the metric can be decomposed as
\beq
\delta g_{ij} = h_{ij}^{\hbox{\scriptsize TT}} + \delta\phi\, g_{ij}
  + (L\xi)_{ij} ,
\label{b1}
\eeq
where
\beq
(L\xi)_{ij} = \nabla_i\xi_j + \nabla_j\xi_i
  - {1\over n} g_{ij}\nabla_k\xi^k ,
\label{b2}
\eeq
and the transverse traceless deformation $h_{ij}^{\hbox{\scriptsize TT}}$
satisfies
\beq
g^{ij}h_{ij}^{\hbox{\scriptsize TT}} = 0 , \qquad
(L^\dagger h^{\hbox{\scriptsize TT}})_i
= -2\nabla^jh_{ij}^{\hbox{\scriptsize TT}} = 0 .
\label{b3}
\eeq
Now observe that the conformal variation $\delta\phi$ is determined
by the Hamiltonian constraint, and that the term $(L\xi)_{ij}$ is, up
to a conformal piece, ``pure gauge''---it is essentially the deformation
${\cal L}_\xi g_{ij}$ of $g_{ij}$ induced by the infinitesimal
diffeomorphism generated by the vector field $\xi^i$.  The true dynamical
degrees of freedom are therefore, at least infinitesimally, the transverse
traceless variations $h_{ij}^{\hbox{\scriptsize TT}}$.  In particular,
when $n=2$, the kernel of $L^\dagger$ is finite dimensional---it is the
space of quadratic differentials \cite{Gardiner}---and the physical
configuration space has only finitely many degrees of freedom.
\vskip 0.2cm
\noindent {\bf 2.\ A constraint analysis} \cite{ADM}: The momentum
constraint of canonical general relativity takes the form
\beq
{\cal H}^i = -2\nabla_j \pi^{ij} = 0 ,
\label{b4}
\eeq
where $\pi^{ij}$ is the momentum conjugate to $g_{ij}$.  The canonical
generator of spatial diffeomorphisms, on the other hand, is
\beq
G[\xi] = \int_\Sigma\! d^nx\, (\nabla_i\xi_j + \nabla_j\xi_i)\pi^{ij}
  = \int_\Sigma\! d^nx\, \xi_i{\cal H}^i .
\label{b5}
\eeq
That is, the standard Poisson brackets of the canonical variables imply
that
\beq
\left\{ G[\xi], g_{ij} \right\} = \nabla_i\xi_j + \nabla_j\xi_i
  = {\cal L}_\xi g_{ij}
\label{b6}
\eeq
along with the corresponding expression for the momentum.  Invariance
under spatial diffeomorphisms thus follows from the vanishing of the
constraints, and variations of the metric of the form \rref{b6} are
thus nonphysical.  A similar argument relates diffeomorphisms generated
by timelike vector fields to the Hamiltonian constraint $\cal H$, at
least on shell.

Note, however, that both of these arguments implicitly assumed that
$\Sigma$ had no boundary.  In the presence of a boundary, the splitting
\rref{b1} is well-defined and unique only if one chooses boundary
conditions that make the operator $L^\dagger L$ self-adjoint.  The
simplest such choice is
\beq
\xi^i \left|_{\partial\Sigma}\right. = 0 ,
\label{b7}
\eeq
which restricts the ``pure gauge'' degrees of freedom to those generated
by vector fields that vanish on the boundary.
Similarly, equation \rref{b5} required an integration by parts; if
$\Sigma$ has a boundary, we find instead
\beq
G[\xi] = \int_\Sigma\! d^nx\, \xi_i{\cal H}^i +
  2\int_{\partial\Sigma}\! d^{n-1}x\,\xi_i\pi^{i\perp} .
\label{b8}
\eeq
The last term in \rref{b8} vanishes only for vector fields satisfying
the boundary conditions \rref{b7}.  For vector fields that do not vanish
on $\partial\Sigma$, the constraints no longer imply invariance, but
merely relate diffeomorphisms to a new set of variables
\beq
{\cal O}[\xi] = \int_{\partial\Sigma}\! d^{n-1}x\, \xi_i\pi^{i\perp}
\label{b9}
\eeq
at the boundary.  Balachandran et al.\ have shown that these are
{\em physical\/} variables, that is, that they commute with the
constraints, thus providing a new set of boundary observables
\cite{Bal1,Bal2}.  In the corresponding quantum theory, we thus
expect a new set of operators associated with the boundary, and a
new set of physical degrees of freedom, the ``would-be pure gauge''
degrees of freedom associated with vector fields that do not vanish
at the boundary.

Additional evidence for new degrees of freedom associated with boundaries
comes from a variety of sources.  For example, Esposito et al.\ have
shown that in the one-loop computation of the partition function for
Euclidean gravity on a four-ball, the gauge and ghost contributions
do not cancel, as they do for a closed manifold, but instead give a
necessary contribution to the path integral \cite{Esposito}.  Baez et
al.\ \cite{Baez} and Smolin \cite{Smolin} have investigated the loop
variable approach to quantum gravity in the presence of boundaries,
and have also found evidence for boundary degrees of freedom.  Finally,
in the nongravitational context it is well known that Chern-Simons theory
on a manifold with boundary induces a dynamical Wess-Zumino-Witten (WZW)
theory on the boundary \cite{Witten,EMSS}, whose degrees of freedom can
be understood as ``would-be gauge'' degrees of freedom of precisely the
kind described above \cite{Ogura,CarWZW,Bal3}.
\vskip 1cm

\noindent {\bf III. The Horizon as a Boundary}
\vskip 0.7cm
Such ``extra'' boundary degrees of freedom are natural candidates
for the microscopic degrees of freedom responsible for black hole
thermodynamics.  In particular, if we treat the horizon of a black
hole as a boundary, we will obtain new states whose number might be
related to the black hole's entropy.  There is an obvious objection
to this viewpoint, however: the event horizon of a black hole is
not, in fact, a boundary.

But while the horizon is not a true boundary of spacetime, it is
a surface upon which we impose boundary conditions.  Any quantum
mechanical statement about black holes is necessarily a statement
about {\em conditional\/} probabilities: for instance, ``If spacetime
contains an event horizon of a certain size, then we should see Hawking
radiation with a certain spectrum.''  To impose the condition
(``spacetime contains an event horizon of a certain size''), we must
fix ``boundary'' data, requiring the existence of a hypersurface with
appropriate geometric properties---vanishing expansion of outgoing null
geodesics, for example.

In a path integral approach to quantization, there is an obvious
way to do this: we can split spacetime $M$ into two pieces, say
$M_1$ and $M_2$, along a hypersurface $\Sigma$, and perform separate
path integrals in $M_1$ and $M_2$ with suitable boundary conditions
at $\Sigma$.  Such a procedure has been studied extensively,
particularly in the context of two-dimensional conformal field theory
\cite{Car3,CCDD,Sonoda}, as has the converse problem of ``sewing,''
that is, integrating over boundary data on $\Sigma$ to recover the
path integral on $M$.  In particular, it is known that in order to
``sew'' properly, the actions on $M_1$ and $M_2$ must sometimes be
supplemented by boundary terms, and these boundary terms may break
gauge invariance and give dynamics to ``pure gauge'' degrees of
freedom at the boundary.

An important example of this phenomenon is Chern-Simons theory.
Consider for simplicity an abelian $\hbox{U}(1)$ Chern-Simons
theory, described by the action
\beq
I_M[A] = {k\over2\pi}
   \int_M \!d^3x\,\epsilon^{\mu\nu\rho}A_\mu\partial_\nu A_\rho ,
\label{c1}
\eeq
where $M$ is a closed three-manifold.  This action is invariant under
gauge transformations
\beq
A_\mu\rightarrow A_\mu+\partial_\mu\Lambda ,
\label{c2}
\eeq
and leads to Euler-Lagrange equations
\beq
F_{\mu\nu} = \partial_\mu A_\nu - \partial_\nu A_\mu = 0 .
\label{c3}
\eeq
The space of classical solutions is thus the space of flat connections
modulo gauge transformations.  The corresponding quantum theory is
fairly simple, and it may be shown to have a finite-dimensional Hilbert
space (see, for example, \cite{Poly}).

Let us now split $M$ into two pieces along a surface $\Sigma$, and
consider the action \rref{c1} restricted to, say, $M_1$.  On a manifold
with boundary, the variation of the action gives
\beq
\delta I_{M_1}[A] = (\hbox{Euler-Lagrange equations})
  - {k\over2\pi}\int_\Sigma \!d^2x\, n_\rho
  \epsilon^{\rho\mu\nu}A_\mu\delta A_\nu ,
\label{c4}
\eeq
and the boundary term in \rref{c4} means that there are typically no
classical extrema.  It is known, at least in many examples, that in order
to ensure proper ``sewing'' of transition amplitudes, we must add boundary
terms to the action in a way that guarantees that extrema exist for a
sufficiently large set of boundary data.  In Chern-Simons theory, the
standard approach is to choose a complex structure on $\Sigma$ and to
fix the boundary value of the component $A_z$, which is canonically
conjugate to $A_{\bar z}$.  The boundary term in \rref{c4} can then be
cancelled by a boundary action
\beq
I_\Sigma[A] = {k\over2\pi}\int_\Sigma \!d^2x\, A_zA_{\bar z} .
\label{c5}
\eeq

Observe now that the action
\beq
I_{M_1}'[A] = I_{M_1}[A] + I_\Sigma[A]
\label{c6}
\eeq
is no longer invariant under the gauge transformations \rref{c2} unless
$\Lambda$ vanishes at the boundary.  This feature should look familiar:
it is a gauge theoretical analog of the gravitational phenomenon we saw
in the preceding section.  We can make this noninvariance explicit by
decomposing $A_\mu$ as
\beq
A_\mu = \bar A_\mu + \partial_\mu\Lambda ,
\label{c7}
\eeq
where $\bar A_\mu$ is a gauge-fixed potential; then
\beq
I_{M_1}'[A] = I_{M_1}'[\bar A]
   + {k\over2\pi}\int_\Sigma \!d^2x
   \left(\partial_z\Lambda\partial_{\bar z}\Lambda
   + 2 \bar A_z\partial_{\bar z}\Lambda \right) .
\label{c8}
\eeq
The would-be gauge transformation $\Lambda$ has thus become a dynamical
field on $\Sigma$, with an action that can be recognized as a chiral
Wess-Zumino-Witten action.  This is a dramatic result: we have gone from
a Chern-Simons theory with a finite-dimensional Hilbert space to a theory
that includes an infinite-dimensional Hilbert space describing boundary
degrees of freedom.

An analogous process occurs in the nonabelian case.  Let $A =
A_\mu^a T_a dx^\mu$ denote a connection one-form for a nonabelian
gauge group $G$ with generators $T_a$.  Then the Chern-Simons action
\beq
I_{M_1}'[A] = {k\over4\pi}\int_{M_1} \hbox{Tr}\left(
   A\wedge dA + {2\over3}A\wedge A\wedge A\right)
   + {k\over4\pi}\int_\Sigma \hbox{Tr}A_zA_{\bar z}
\label{c9}
\eeq
appropriate for fixing $A_z$ at $\Sigma$ again splits into two pieces;
under the decomposition
\beq
A = g^{-1}dg + g^{-1}\bar A g ,
\label{c10}
\eeq
the action becomes \cite{Ogura,CarWZW}
\beq
I_{M_1}'[\bar A,g]
   = I_{M_1}'[\bar A] + k I^+_{\hbox{\scriptsize WZW}}[g,\bar A_z] ,
\label{c11}
\eeq
where $I^+_{\hbox{\scriptsize WZW}}[g,\bar A_z]$ is now the action of
a nonabelian chiral WZW model on the boundary $\Sigma$,
\beq
I^+_{\hbox{\scriptsize WZW}}[g,\bar A_z]
 = {1\over4\pi}\int_\Sigma\hbox{Tr}
 \left(g^{-1}\partial_z g\,g^{-1}\partial_{\bar z} g
 - 2g^{-1}\partial_{\bar z} g {\bar A}_z\right)
 + {1\over12\pi}\int_M\hbox{Tr}\left(g^{-1}dg\right)^3 .
\label{c12}
\eeq
Witten has shown that when this WZW action is included, Chern-Simons
theory ``sews'' properly at the boundary $\Sigma$ \cite{WitWZW}.
\vskip 1cm

\noindent {\bf IV. A (2+1)-Dimensional Model}
\vskip .7cm
The discussion has so far been rather general.  We have seen that there
are, plausibly, new degrees of freedom associated with black hole horizons;
and we have seen that for a particular theory that is {\em not\/} gravity,
similar degrees of freedom lead to interesting dynamics.  The next step
should be to put these two ingedients together to find the dynamics of
our gravitational boundary degrees of freedom.

Unfortunately, this is not easy.  The basic difficulty can be seen by
comparing equations \rref{b1} and \rref{c10}.  For the gauge theory,
the splitting of $A$ into ``physical'' and ``gauge'' degrees of freedom
is a local decomposition, valid for arbitrary finite gauge transformations.
For gravity, on the other hand, the decomposition has only been written
infinitesimally; because diffeomorphisms move points, the analog of
\rref{c10} for finite diffeomorphisms is highly nonlocal.

In 2+1 dimensions, this difficulty can be avoided.  As first shown by
Ach{\'u}carro and Townsend \cite{Achu}, general relativity in three
spacetime dimensions can be rewritten as a Chern-Simons theory, with
diffeomorphisms replaced by ordinary gauge transformations.  In
particular, if the cosmological constant is negative---as required
for the existence of a black hole \cite{BTZ}---then general relativity
is equivalent to an $\hbox{SO}(2,1)\!\times\!\hbox{SO}(2,1)$
Chern-Simons theory.  As is the case for other Chern-Simons theories,
(2+1)-dimensional gravity therefore induces a Wess-Zumino-Witten action
on spatial boundaries, and we can study the dynamics of the ``would-be
pure gauge'' degrees of freedom in some detail.

The results of such an analysis can be summarized as follows \cite{Car2}:
\begin{enumerate}
\item The diffeomorphisms and local Lorentz transformations of general
relativity are equivalent to $\hbox{SO}(2,1)\!\times\!\hbox{SO}(2,1)$
gauge transformations of the Chern-Simons theory \cite{Witten2}.
\item On a manifold with boundary, (2+1)-dimensional gravity in its
Chern-Simons form induces an $\hbox{SO}(2,1)\!\times\!\hbox{SO}(2,1)$
WZW action on the boundary.  With some reasonable assumptions about
quantization, the states of this boundary theory can be written down
explicitly.
\item While almost all of the diffeomorphisms of the boundary become
dynamical, the theory retains a remnant of diffeomorphism invariance:
the diffeomorphisms generated by Killing vectors---which are in the kernel
of $L$ and thus missing from equation \rref{b1}---generate invariances
that must be respected by the physical states.
\item When this physical state condition is imposed, the number of
states is finite, and can be estimated by standard number theoretical
arguments.  Given a fixed horizon size, the resulting number of states
is the exponential of the correct Bekenstein-Hawking entropy,
\beq
n(r_+) = \exp\left\{ {2\pi r_+\over 4\hbar G}\right\} .
\label{d1}
\eeq
\end{enumerate}

Let me now give a few more details.  (For a full description, see
\cite{Car2}.)  If we write the cosmological constant as $\Lambda=-1/\ell^2$,
we can define two $\hbox{SO}(2,1)$ gauge fields,
\beq
A^a = \omega^a + {1\over\ell}e^a , \quad
      \tilde A^a = \omega^a - {1\over\ell}e^a .
\label{d2}
\eeq
Here $e^a = e^a_\mu dx^\mu$ is a triad, $\omega^a = {1\over2}
\epsilon^{abc}\omega_{\mu bc}dx^\mu$ is a spin connection, and
\beq
k = {\ell\sqrt{2}\over 8G}
\label{d3}
\eeq
in the normalizations of reference \cite{Car2}.  The Einstein-Hilbert
action of general relativity is then
\beq
I_{\hbox{\scriptsize grav}}
  = I_{\hbox{\scriptsize CS}}[A] - I_{\hbox{\scriptsize CS}}[\tilde A] ,
\label{d4}
\eeq
where $I_{\hbox{\scriptsize CS}}[A]$ is the Chern-Simons action \rref{c9}.
As described above, this action must be supplemented with appropriate
boundary terms if the horizon is treated as a boundary.  The nature of
these terms depends on the choice of boundary conditions; for a black
hole, we can demand that $\partial M$ be a null surface and that the
expansion $\theta^+$ of outgoing null geodesics vanish, so that
$\partial M$ is an apparent horizon.  The resulting action induces
an $\hbox{SO}(2,1)\!\times\!\hbox{SO}(2,1)$ chiral WZW theory on
$\partial M$ in a manner exactly analogous to the appearance of the
abelian WZW action in \rref{c8}.

This boundary action is completely characterized by a current algebra
\cite{GepWit}
\beq
[J^a_m,J^b_n] = if^{ab}{}_cJ^c_{m+n} - km{\hat g}^{ab}\delta_{m+n,0} ,
  \quad [\tilde J^a_m,\tilde J^b_n] = if^{ab}{}_c\tilde J^c_{m+n}
  + km{\hat g}^{ab}\delta_{m+n,0} ,
\label{d5}
\eeq
where $\hat g$ is the Cartan-Killing metric, and the zero-modes $J^a_0$
and $\tilde J^a_0$ of the currents are fixed by the boundary data on
$\partial M$.  Because the group is noncompact, the quantization of
the $\hbox{SO}(2,1)$ WZW model is not completely understood, but in the
large $k$---i.e., small $\Lambda$---limit, the model may be approximated
by a theory of six independent bosonic string oscillators.  Such a system
has an infinite number of states, which can be generated from a vacuum
$|0\rangle$ that satisfies
\beq
J^a_n |0\rangle = \tilde J^a_n |0\rangle = 0 \quad\hbox{for $n>0$}
\label{d6}
\eeq
by acting with $J^a_{-n}$ and $\tilde J^a_{-n}$.

We have not yet applied the physical state condition, however.  Recall
that the horizon fields can be interpreted as ``would-be pure gauge''
excitations that become dynamical at a boundary.  It is apparent from
equation \rref{b1} (or equations \rref{b5} and \rref{b9}), however, that
the excitations corresponding to Killing vectors remain genuine gauge
degrees of freedom.  We must therefore impose a remaining Wheeler-DeWitt
equation, requiring that physical states be invariant under those
diffeomorphisms that are generated by Killing vectors.  For the
(2+1)-dimensional black hole background, this requirement is that
\beq
{\hat L}_0 |\hbox{phys}\rangle = 0 ,
\label{d7}
\eeq
where ${\hat L}_0$ is the zero-mode of the Virasoro generator associated
with the affine algebra \rref{d5}.

Now, ${\hat L}_0$ has a standard expression in terms of the currents
\rref{d5}:
\beq
{\hat L}_0 = \sum_{i=1}^6 N_i + (\hbox{current zero-mode pieces}) ,
\label{d8}
\eeq
where the $N_i$ are number operators (six because there are three
components of $J^a$ and three components of $\tilde J^a$).  Imposing
\rref{d7} thus fixes this sum of number operators in terms of zero-modes,
which are in turn determined by boundary data at the black hole horizon.
After a bit of manipulation, we find the condition
\beq
\sum_{i=1}^6 N_i = \left({r_+\over4G}\right)^2 ,
\label{d9}
\eeq
where $r_+$ is the horizon radius.

For large values of $r_+$, the number of states satisfying this condition
can be found by a number theoretical argument that dates back to Ramanujan
and Hardy \cite{Raman}.  The result is equation \rref{d1}.  But the factor
in the exponent in \rref{d1} is precisely the correct Bekenstein-Hawking
entropy for a (2+1)-dimensional black hole \cite{CarTeit}, so our
``would-be pure gauge'' degrees of freedom do, in fact, explain the
entropy.

While this argument is quite convincing, it is not entirely satisfactory,
due to limitations in our current understanding of WZW models for
noncompact groups.  Witten has suggested a more rigorous approach to the
quantization of Euclidean gravity in 2+1 dimensions, which leads naturally
to a connection with $\hbox{SU}(2)$ WZW models \cite{Witten3}.  It should
be possible to investigate the entropy of the (2+1)-dimensional black
hole in this context; preliminary results indicate that the expression
\rref{d1} for the number of states can be reproduced.
\vskip 1cm

\noindent {\bf V. Conclusion}
\vskip .7cm
These results for the entropy of the (2+1)-dimensional black hole are
exciting, but they are also frustrating.  The methods described here
do not generalize to 3+1 dimensions, and while we may argue that the
horizon degrees of freedom still exist, it is not clear how to count
them.  A useful step in this direction would be to find the appropriate
symplectic structure for the ``would-be pure gauge'' degrees of freedom
on the horizon; Epp has recently made some progress in this direction
\cite{Epp}.  We must also understand the appropriate boundary terms at
the horizon in the conventional metric formulation of general relativity,
a problem for which recent results of Teitelboim should be relevant
\cite{Teitel}.  It would also be interesting to look at the abstract
quantization of the group $\hbox{\em Diff}\,S^2$ of diffeomorphisms of
the two-sphere, which might relate to our problem in the same way
coadjoint orbit quantization of $\hbox{\em Diff}\,S^1$ relates to WZW
theory \cite{coadjoint}.

Finally, it is interesting to note that the {\em techniques\/} used to
count horizon states in 2+1 dimensions are remarkably similar to the
methods that string theorists have recently used to determine black hole
entropy \cite{Horowitz}.  The physical starting points seem very different,
but the appearance of such similar mathematics suggests that there may be
hidden connections.
\vskip 1cm

\newpage
\noindent {\bf Acknowledgements}
\vskip .7cm
This work was supported in part by U.S.\ National Science Foundation grant
PHY-93-57203 and U.S.\ Department of Energy grant DE-FG03-91ER40674.

\renewcommand{\refname}
\end{document}